\documentclass[prl,preprint,footinbib,showpacs]{revtex4}

\usepackage[utf8]{inputenc}

\usepackage{amssymb}
\usepackage{amsmath}
\usepackage{amsbsy}
\usepackage{bm}
\usepackage{graphicx}

\newcommand{\lz}{l_0}
\newcommand{\tz}{t_0}

\newcommand{\tfmu}{\mathcal{F}^{-1}}
\newcommand{\vk}{\mathbf{k}}

\newcommand{\vecr}{\mathbf{r}}

\newcommand{\nf}{\nu_f}

\newcommand{\Ef}{E_f}

\newcommand{\cEz}{\mathcal{E}^0}
\newcommand{\pe}{\! = \!}

\newcommand{\ppp}{\! < \!}
\newcommand{\ppg}{\! > \!}

\newcommand{\hti}{\hat{h}}
\newcommand{\tb}{t}

\newcommand{\cw}{c_w}
\newcommand{\calH}{\mathcal{H}}
\newcommand{\cF}{\mathcal{F}}

\newcommand{\fud}{\frac{1}{2}}

\newcommand{\Fel}{\mathcal{F}^{\mbox{\scriptsize{{\rm{El}}}}}}
\newcommand{\muel}{\mu^{\mbox{\scriptsize{{\rm{El}}}}}}
\newcommand{\Fsurf}{\mathcal{F}^{S}}

\newcommand{\hz}{\bar{h}}
\newcommand{\hs}{h_s}
\newcommand{\Fa}{\tilde{F}}
\newcommand{\deltaa}{\delta}
\newcommand{\kb}{k_{{\scriptscriptstyle{B}}}}
\newcommand{\cV}{\mathcal{V}}
\newcommand{\hdep}{h_{d}}

\begin{document}

\title{Dynamics and ordering of the nucleationless island formation in heteroepitaxy}

\author{Jean-No\"el Aqua$^{1,2}$}
\altaffiliation{\'Ecole Centrale Marseille}
\affiliation{$^1$ Institut Mat\'eriaux Micro\'electronique
Nanosciences de Provence, Aix-Marseille Universit\'e,  UMR 6242, 13397 Marseille, France\\
$^2$ Institut de Recherche sur les Ph\'enom\`enes Hors
\'Equilibre,  Aix-Marseille Universit\'e, UMR 6594, 13384 Marseille, France
}
\author{Thomas Frisch$^{1}$} 
\altaffiliation{\'Ecole Centrale Marseille}
\affiliation{$^1$ Institut Mat\'eriaux Micro\'electronique
Nanosciences de Provence, Aix-Marseille Universit\'e,  UMR 6242, 13397 Marseille, France\\
$^2$ Institut de Recherche sur les Ph\'enom\`enes Hors
\'Equilibre,  Aix-Marseille Universit\'e, UMR 6594, 13384 Marseille, France
}

\author{Alberto Verga$^1$}
\affiliation{$^1$ Institut Mat\'eriaux Micro\'electronique
Nanosciences de Provence, Aix-Marseille Universit\'e,  UMR 6242, 13397 Marseille, France\\
$^2$ Institut de Recherche sur les Ph\'enom\`enes Hors
\'Equilibre,  Aix-Marseille Universit\'e, UMR 6594, 13384 Marseille, France
}

\date{\today}

\begin{abstract}
We study the morphological evolution of strained islands in growing crystal films by 
use of a continuum description including wetting, elasticity and deposition flux. 
Wetting breaks translational invariance, allowing the flux to tune different 
nonlinear regimes. Increasing the flux, we find first an annealing-like dynamics, then a slower but 
non-conventional ripening followed by a steady regime, while the island density continuously increases. 
The islands develop spatial correlations and ordering with a narrow two-peaked distance distribution and
 ridge-like clusters of islands for high flux. 
\end{abstract}

\maketitle


Non-equilibrium crystal growth is questioning many fundamental and experimental 
issues in particular in the domain of self-organization of nanostructures
\cite{Shchukin,Stangl04,Teichert02,Pimpinelli98,BaraStan}. For example, 
quantum dots has lead to numerous applications \cite{Vved08} such as photovoltaic cells, 
memory storage or light emission. However, 
the different scenario governing island formation are still challenged
as regards their density, size distribution and spatial ordering. 
We focus here on the properties of islands
produced in heteroepitaxy.

When a film is coherently deposited on a substrate with a lattice mismatch, it experiences 
an elastic stress that can be relieved by a morphological evolution. For strong enough mismatch,  
islands are nucleated in an abrupt two to three dimensions transition. However, 
for intermediate mismatch, the evolution begins by surface diffusion with a nucleationless ripple formation 
\cite{SuttLaga00,TromRoss00,BerbRond02,Flor00} which results from 
an elastic instability reminiscent of the Asaro-Tiller-Grinfel'd 
instability \cite{GaoNix99,SpenVoor91}. Contrarily to the evolution in thick films
 \cite{GaoNix99}, no dislocations are generated
in thin films where instead the ripples transform smoothly into islands separated by a wetting layer
\cite{BerbRond02}. This scenario requires full understanding of the nonlinear regime which involves 
nonlocal elastic  interactions \cite{Spencer93,XianE02,Golovin03,jnFrisVerg07,Levine07}. 
Crucial for possible application, the resulting spatial order 
depends strongly on the growth dynamics. 
Hence, the question we address in this Letter is the influence of the dynamics 
on the spatial organization of the islands. 
To tackle this problem, we use a continuum modelization of the crystal accounting both 
for wetting and elastic interactions. Wetting, which breaks the translational 
invariance in the growth direction, introduces a significant flux dependence. 
Consequently, we find different nonlinear regimes as the flux $F$ increases. Departing from the near 
annealing case, island ordering arises first as clusters form. Then, for high enough fluxes,
the island density is frozen and the dot density and ordering are maximum. 


The dynamics of a surface during crystal growth involves different mechanisms
such as diffusion, attachment or relaxation. The evolution of its interface 
$z\pe H(\vecr,t)$ at position $\vecr \pe (x,y)$ with time $t$ can be written as
\begin{equation}
 \label{surfevol}
    \frac{\partial H}{\partial t} = \cV[H] + a_f  F , 
\end{equation}
where $\cV[H]$ is dictated by the predominant mechanisms at stake while  
$a_f$ is the film lattice parameter. 
In homoepitaxy, symmetry constraints enforce $\cV$ to depend only 
on the slope of $H$ \cite{BaraStan} so that $F$ disappears in the Galilean transformation 
$H(\vecr,t)= F t + h (\vecr,t)$. 
This invariance is violated in heteroepitaxy when a film is coherently 
deposited on a substrate. We consider a film evolving in the Stransky-Krastanov mode, with 
a dynamics due to surface diffusion induced by chemical potential gradients. Hence, 
\begin{equation}
\cV [H] = D \sqrt{1+|\nabla H|^2} \, \nabla_S^2 \mu ,
\end{equation}
where $D$ is a diffusion coefficient, $\nabla_S$, the surface gradient, and $\mu$, 
the surface chemical potential given by the functional derivative 
$\mu \pe \delta (\Fel + \Fsurf)/\delta H$, where $\Fsurf \pe \int d\vecr 
\gamma(H) \sqrt{1+|\nabla H|^2}$ and $\Fel$ is the elastic free energy. 
Wetting is embedded in the $H$-dependence of the surface energy $\gamma$, which 
describes the change with $H$ of the local environment of a particle when the film/substrate 
interface is present \cite{Spen99}, 
and which precisely breaks translational invariance in the $z$ direction. 
Note that neither alloying nor anisotropy are considered here, which can prove significant in some 
systems \cite{TuTers04,TersSpen02}. 

To fix scales, we choose to depict a Ge$_{0.75}$Si$_{0.25}$-like film
deposited on a Si substrate with a reference lattice with a substrate (film) lattice 
parameter $a_s \pe 0.27$\,nm ($a_f \pe 1.01 a_s$). Surface diffusion is given by 
$D\pe D_s\exp[-E_d/\kb T] a_f^4 / \kb T$ 
with $E_d \pe 0.83$\,eV and $D_s\pe 8.45 \, 10^{-10}$\,m$^2$/s, see e.g.~\cite{SpenVoor91}, 
and the working temperature is $700^{\,\textrm{o}}$C. The film surface energy is 
$\gamma_f \pe 1.3$\,J/m$^2$ and we extrapolate ab-initio calculations for Si/Ge systems \cite{abinitio} 
by considering 
$\gamma(h) \pe \gamma_f  [ 1+ \cw \exp(-h/\delta)]$ where $\delta\pe a_f$ and $\cw \pe 0.09$. 
The characteristic length and time scales are then 
$\lz \pe \cEz / 2(1+\nf) \gamma_f$ and $\tz \pe \lz^4/D \gamma_f$, 
where $\cEz \pe \Ef \, (a_f-a_s)^2/a_s^2 (1-\nf)$
is an elastic energy density involving the film Poisson ratio and Young modulus $\nf$ and 
$\Ef$, whose values are $\lz \pe 27$\,nm and $\tz \pe 25$\,s.

Mechanical equilibrium is supposed to be achieved on time scales faster than the system evolution, 
which enforce the Lamé equations $\partial_q \sigma_{pq} \pe 0$ to be valid
in the film and substrate. The stress tensor $\sigma_{pq}$ is a linear function of the 
strain tensor  $e_{pq} \pe \fud (\partial_q u_p + \partial_p u_q) - e_{pq}^r$ where $\mathbf{u}$ 
is the displacement
with respect to the substrate reference state. The reference strain is 
$e_{pq}^r \pe (1-a_f/a_s) \, \delta_{pq} (\delta_{px}+\delta_{qy}) $ in the film and $0$ otherwise, 
where $\delta_{p,q}$ is the Kronecker symbol with $p,q\pe x,y,z$. 
To simplify calculations, we consider isotropic elasticity and identical film and substrate 
elastic constants as coherent epitaxy involves similar materials. This approximation neglects 
higher order terms and ensures the translational invariance with $z$ of the elastic energy; 
consequently, in the Galilean frame $h\pe H - \hz(t)$, 
the dependence on $\hz (t)$ will appear exclusively through the wetting term. 
The Lamé equations are then solved thanks to Fourier transforms 
with respect to $\vecr$ \cite{jnFrisVerg07} with the 
following boundary conditions~: the film/substrate interface is coherent with continuous stresses, 
and the film/vacuum surface is free, with a negligible 
surface stress. To solve the latter condition, we use here the 
small-slope approximation amenable for arbitrary deposited thicknesses, 
contrarily to the thin film approximation used in \cite{jnFrisVerg07}. 
Writing $\hz$ the spatial average of $h$, we suppose a shallow modulation 
where $h-\hz$ is small compared to the lateral characteristic length of order $\lz$, and  
get displacements and $\muel$, the elastic energy density 
$\fud \sigma_{pq} e_{pq}$ computed at the film surface, 
up to second order in this small parameter. 
Considering $\hz \pe a_f F t$, the second term in the r.h.s.~of Eq.~\eqref{surfevol} cancels out, 
so that 
\begin{multline}
\label{dhadt}
  \frac{\partial h}{\partial \tb} =   \Delta
  \left\{ \rule{0mm}{5mm} \right.
    	- \left( 1 + \cw e^{-(h+ \Fa t)/\deltaa} \right) \Delta h
    	- \frac{\cw}{\deltaa} \frac{e^{- (h+ \Fa t)/\deltaa}}{\sqrt{1+|\nabla h|^2}} 
    \\
        -  \calH_{ii} (h)
  	    + 2h\Delta h+ \left| \nabla h\right|^2 
    \\
    \left.
    +   2 \calH_{ij}\left[ h\, \theta_{ijkl} \calH_{kl}(h)\right]
         +\calH_{ij}(h) \theta_{ijkl} \calH_{kl}(h)
    	\rule{0mm}{5mm}  \hspace{-0.2mm} \right\},
\end{multline}
in units of $\lz$ and $\tz$ and with  
$\Fa \pe a_f F \tz / \lz$. The long-range elastic interactions 
enforce the non-analytic behavior of the operator
\begin{equation}
\calH_{ij}[h] \pe \tfmu \{ (k_i k_j/|\vk|) \, \cF[h] \} \, ,
\end{equation} 
defined with Fourier transforms $\cF$ over $\vecr$ and wavevector $\vk$,   
where the indices run over $x,y$. The nonlinear terms are given with 
$\theta_{ijij} \pe 1$, $\theta_{iijj} \pe -\theta_{ijji} \pe \nf$ when $i \! \neq \! j$ and 
$0$ otherwise. It is worthwhile to realize that a lone term $h \Delta h$ in the r.h.s. 
of \eqref{dhadt} would be ruled out by symmetry considerations 
\cite{BaraStan} but is allowed here when put in balance with the nonlocal 
nonlinear terms. Indeed, taking into account both local and nonlocal terms ensure the invariance of 
the elastic energy under the transformation $h \! \rightarrow \! h + \hz(t)$ for arbitrary $\hz (t)$.

\begin{figure} \centering
\includegraphics[width=0.43\textwidth]{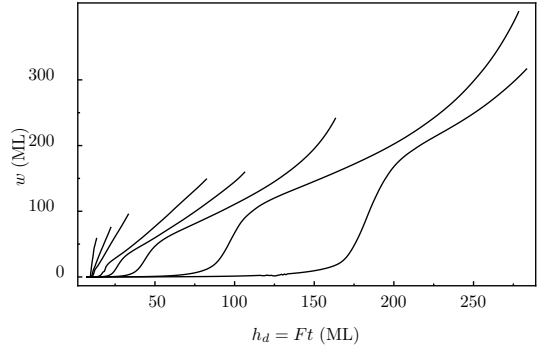}
\caption{Simulation results of the roughness evolution with the deposited height for, 
from left to right, 
$F \pe 10^{-4}, 5\,10^{-4}, 10^{-3}, 5\,10^{-3}, 10^{-2}, 2\,10^{-2}$, 
$5\,10^{-2}, 10^{-1}$\,ML/s. 
} \label{figw2h}
\end{figure}

We performed numerical simulations of Eq.~\eqref{dhadt} using a pseudo-spectral method. 
We consider an initial roughness of amplitude $1$ monolayer (ML) given by a random profile.
The initial film height is 
$\hz_0(t\pe 0)\pe 7$\,ML just below the elastic instability threshold 
$h_c \! \simeq \! 8.2$\,ML defined below. 
Similarly to the annealing case \cite{jnFrisVerg07}, the simulations reveal that the  
combination of wetting and nonlinear nonlocal elastic terms prevents 
the finite-time singularities, which in thick films, lead to dislocations \cite{GaoNix99}. 
In addition, we find that the system evolution depends strongly on the deposition flux.  
Different curves of the roughness $w \pe [\left< h^2 \right> - \left< h \right>^2]^{1/2}$
as function of the deposited height $\hdep \pe F t$ are depicted in Fig.~\ref{figw2h} for different fluxes. 
The roughness first increases exponentially in a linear-like dynamics. After a first inflexion point, the 
system enters a first nonlinear stage where $w \! \sim \! t$, and after a second one, 
displays a faster than linear roughness increase. The deposited heights corresponding to these inflexions 
increase linearly with the flux, though the different curves cannot be rescaled on a single one, signaling 
different ripening mechanisms depending on the flux. 

To evaluate the time of emergence of the islands, we define a dynamical critical 
height $h_s(F)$ after which $w$ is greater than $w_s\pe 3$\,ML. 
It is well fitted by an affine law $h_s \pe 10 \! + \! 1200 F$\,(ML) with a limit at low flux greater than 
$h_c$ due to the threshold present in the definition of $h_s$. 
The values of $h_s$ obtained here for $T \pe 700^{\textrm{o}}$\,C differ from the 
smaller values of the apparent critical thickness derived with the analysis of 
\cite{SpenVoor91}. The latter is defined via a 
comparison of relative growth rates, whereas $h_s$ is defined here directly from the 
roughness which is independent of the reference frame. 
The dynamical critical height $h_s$ is related to the growth in the linear regime and can be roughly 
estimated. In the linear approximation, $h$ small, the evolution of \eqref{dhadt}  can be reformulated 
in Fourier space as
\begin{equation}
\hti (\vk;t) = \hti (\vk;t_0) \, \exp 
 \bigl[ 
  \int_{t_0}^t ds \, \sigma \left(\vk; \hz _0 + \Fa s \right) 
 \bigr] ,
\end{equation}
where
\begin{equation}
\sigma (\vk; \hz) = - k^2 \cw e^{-\hz/\delta}/\delta^2 + |k|^3 - k^4 ( 1+\cw e^{- \hz/\delta} ) .
\end{equation}
By definition of $h_c$, $\sigma$ is negative for $\hz \ppp h_c$ and otherwise
positive in a $\hz$-dependent $k$-interval. To estimate $h_s$, we consider an 
initial undulation with wavevector $k_*$ and an initial amplitude $h_{1,0} \pe 1$\,ML. 
We can solve exactly for $h_1^*(\vecr;t)$ and the critical deposited height and find 
$h_s^* (F) = \xi+\delta \, W(-3 c_w (1+k_*^2 \delta^2) \exp(-\xi/\delta) / k_*^2 \delta^2)$,
with the product-log function $W$ and the length  
$\xi \pe \hz_0 + 4 c_w (1+k_*^2\delta^2)e^{-\hz_0/\delta}/k_*\delta + F \log(w_s/h_{1,0})/k_*^4$.
Considering $k_* \pe 0.35$, the solution is well fitted by $\hs^* (F) \pe 10 \! +\! 1300 F$\,(ML)
which is a rough estimate of $h_s$ for the full equation at low flux. Similarly to the numerical results, 
$h_s^*$ increases nearly linearly with $F$ in the regime of parameters studied here.  
Also, the limit at low fluxes is greater than the instability critical height $h_c$ due to the threshold in the 
definition of $h_s$.  These results could be confronted to experiments investigating
the appearance of islands as function of the ratio $D/F$. 
\begin{figure} \centering
\includegraphics[width=0.12\textwidth]{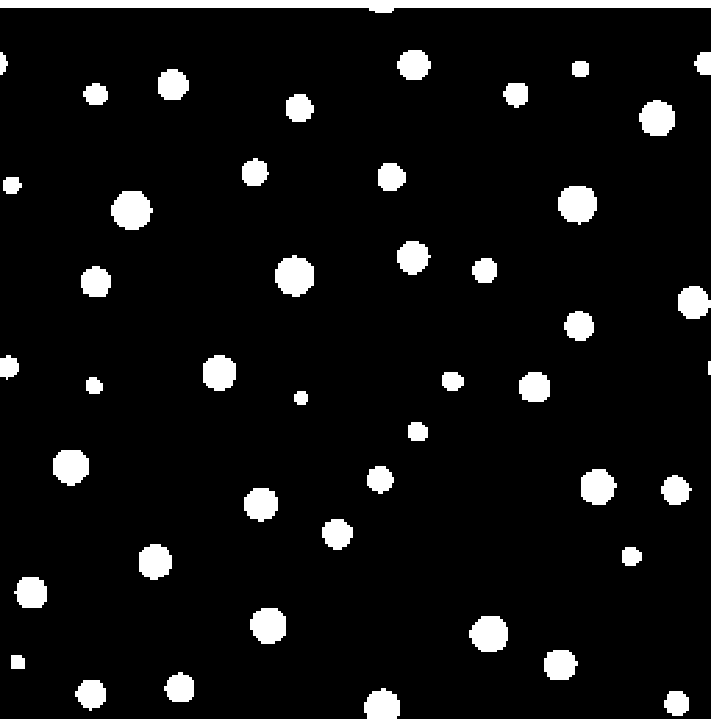}
(a) \hspace{0mm}
\includegraphics[width=0.12\textwidth]{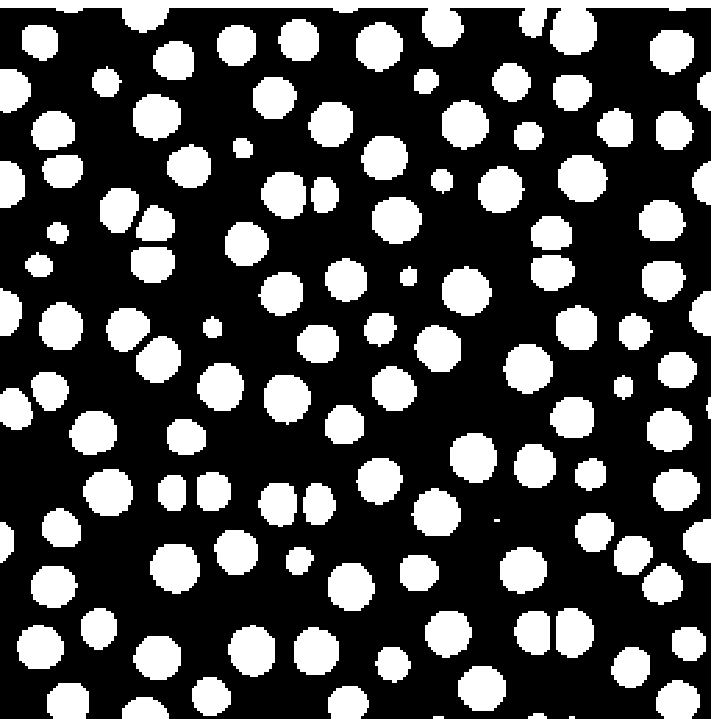}
(b) \hspace{0mm}
\includegraphics[width=0.12\textwidth]{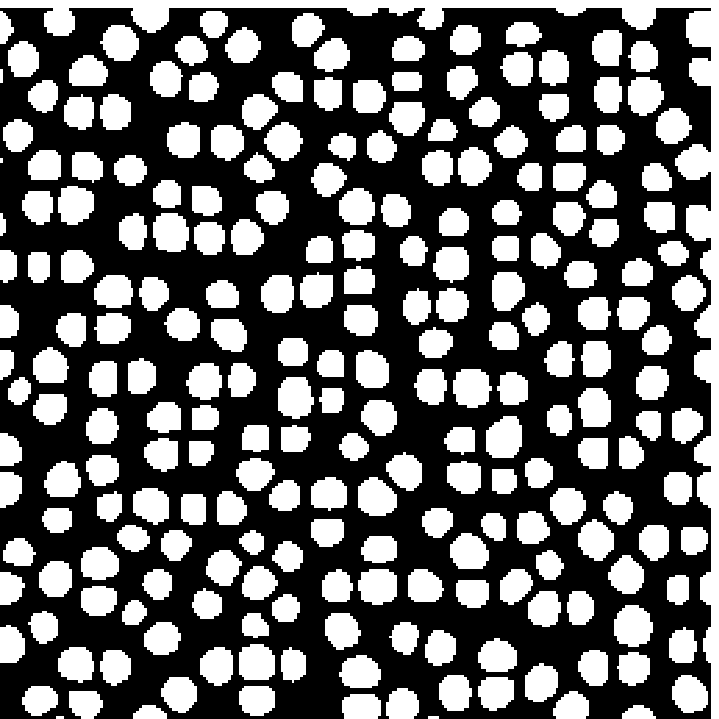}
(c)
\caption{Typical island configurations in the nonlinear regime 
for $F\pe$ (a) $10^{-4}$, (b) $10^{-2}$, (c) $10^{-1}$\,ML/s in a $128 \,\lz \!\times\! 128\,\lz$ system.} \label{figislands}
\end{figure}

\begin{figure} \centering
\includegraphics[width=0.4\textwidth]{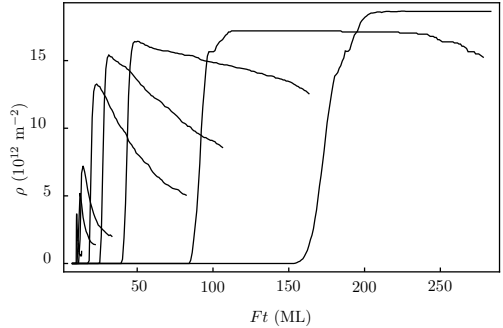}
\caption{Island density as function of the deposited height with, from left to 
right, the fluxes of Fig.~\ref{figw2h}.} \label{figrho2h}
\end{figure}

Above the critical height, the system enters the nonlinear regime characterized 
by islands surrounded by a wetting layer growing both by deposition and coarsening, 
see Fig.~\ref{figislands}. The island density $\rho$ is then plotted in Fig.~\ref{figrho2h} 
for different fluxes. The islands appear all the later that the flux is high, 
as a result of the competition between the instability time scale and the deposition growth. 
In addition, the dynamics depends strongly on the flux. 
At low flux, the island density is convex, similarly to the annealing case, when the system 
has time to coarsen by surface diffusion. However, in the intermediate regime 
for $F$ in between $F_1$ and $F_2$ with $F_1 \! \simeq \! 10^{-2}$\,ML/s and 
$F_2 \! \simeq \!  5\, 10^{-2}$\,ML/s, 
the density evolution becomes concave, whereas in the steady regime, $F \ppg F_2$, 
the island density is constant in a large time interval. The values of $F_1$ and $F_2$ 
depend on the temperature and the details of the wetting that are set here with typical coefficients. 
In all regimes, the island density, together with their areal coverage, at a given deposited height, is an 
increasing function of the flux and saturates at a value limited  by the linear 
instability wavelength. This observation indicates a route for controlling the island 
density. We find also that, in the intermediate and steady 
regimes, the areal coverage first increases with the deposited height even in the nonlinear stage. 

The change of dynamics from a convex island density evolution in the annealing-like case 
to a concave one for growing films is similar to the results of the
experiments by Floro and co-workers \cite{Flor00} where the negativeness of 
$d^2 \rho / d t^2$ is signaling a non standard ripening. Indeed, a typical 
coarsening dynamics tends to decrease its driving force 
resulting in a damped evolution. A faster dynamics can 
nevertheless be apprehended within a mean-field approach describing the island
size distribution evolution due to chemical potentials accounting for 
elastic interactions \cite{Flor00}.  These long range interactions are precisely
the central ingredient of \eqref{dhadt} which can lead to such a dynamics 
when coupled with deposition growth and our simulations exhibit a regime where 
this concave evolution is to be expected. Moreover, in all regimes,  
the dot density evolution is slowed down even when it is given as a 
function of the instability time scale instead of the deposited height. Hence, 
even if a wetting layer has developed and allows surface diffusion,
the stabilization of the islands results from an effective weakening of surface diffusion 
currents when deposition is present. 

\begin{figure} \centering
\includegraphics[width=0.13\textwidth]{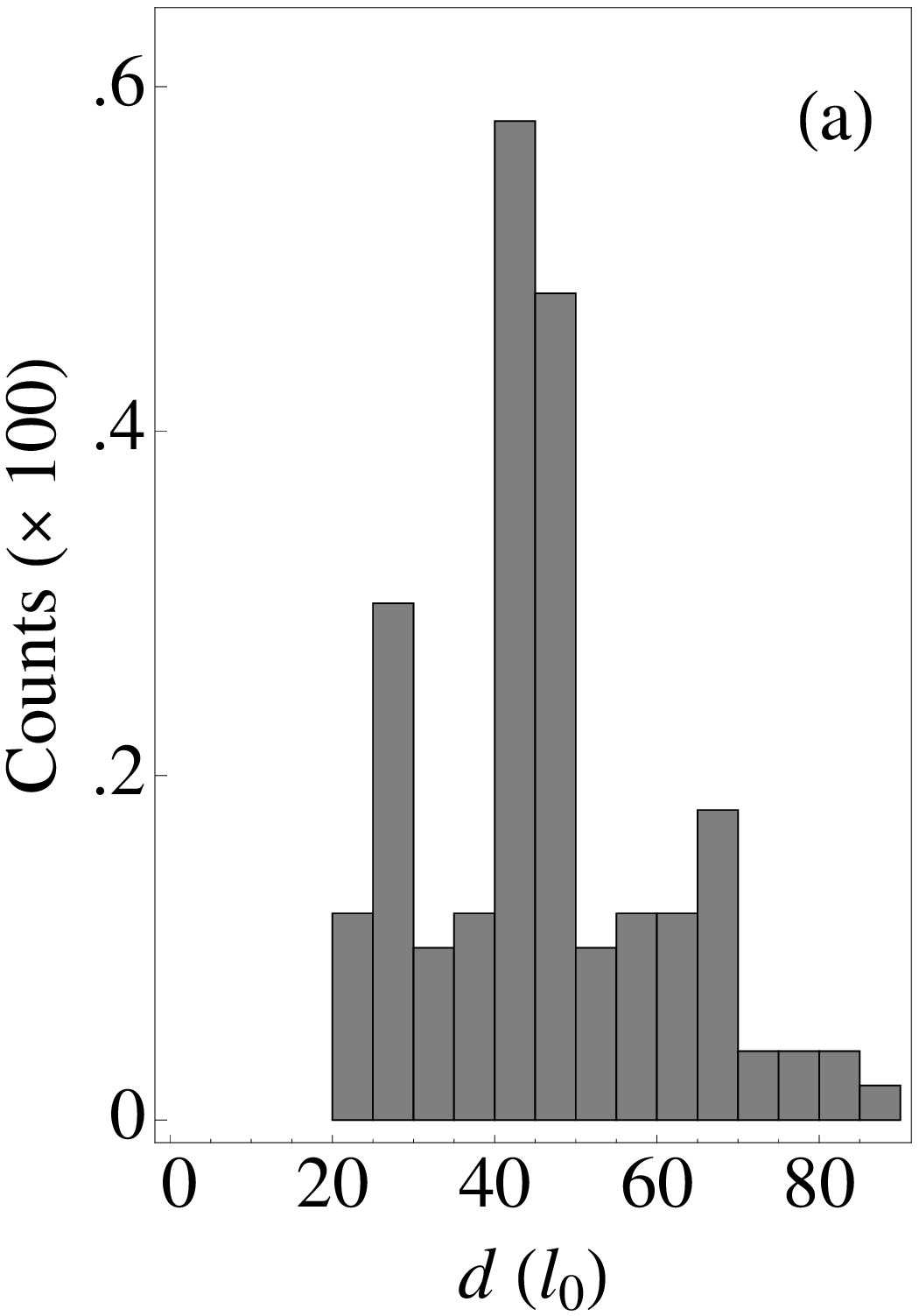} 
\includegraphics[width=0.13\textwidth]{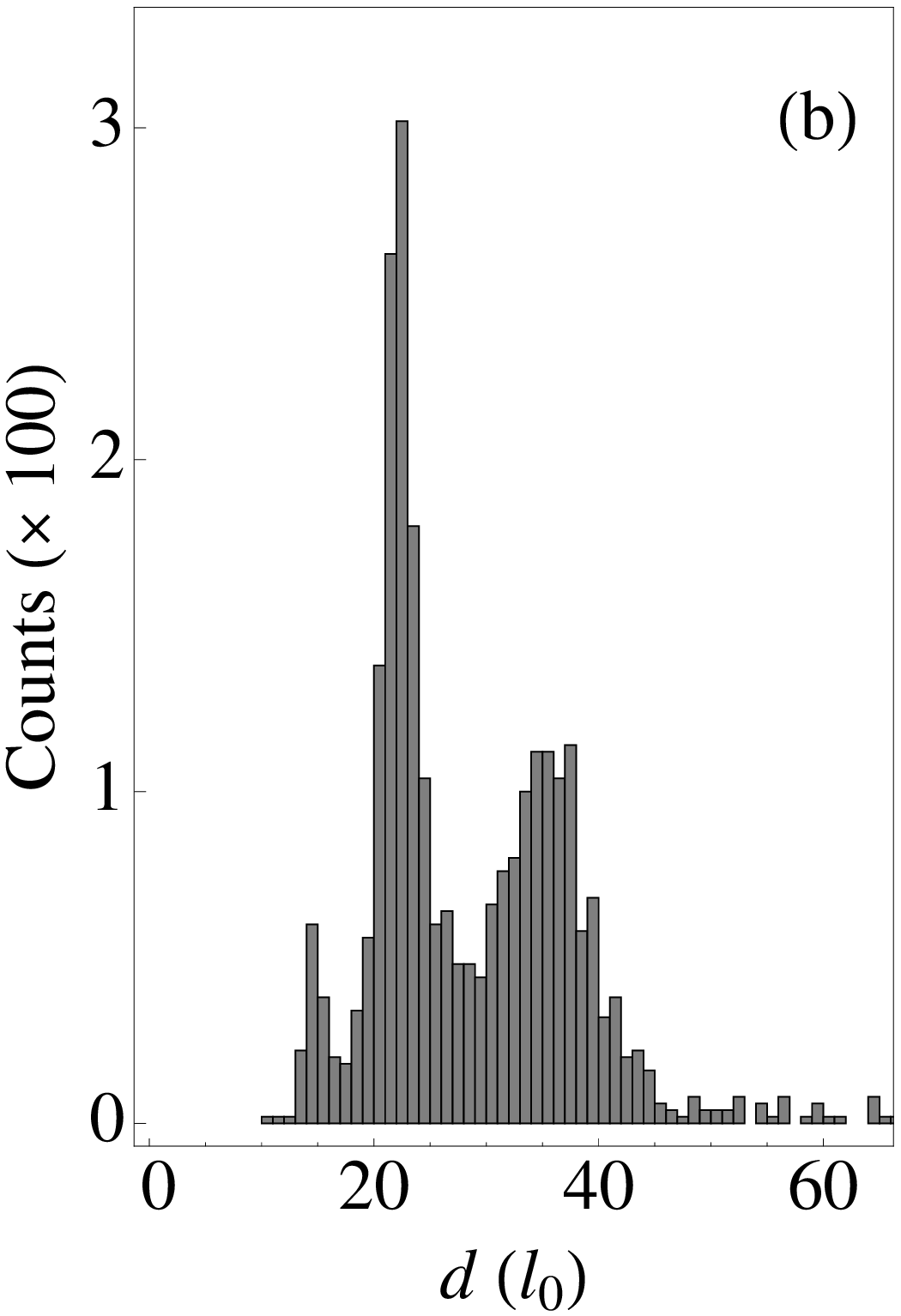}
\includegraphics[width=0.13\textwidth]{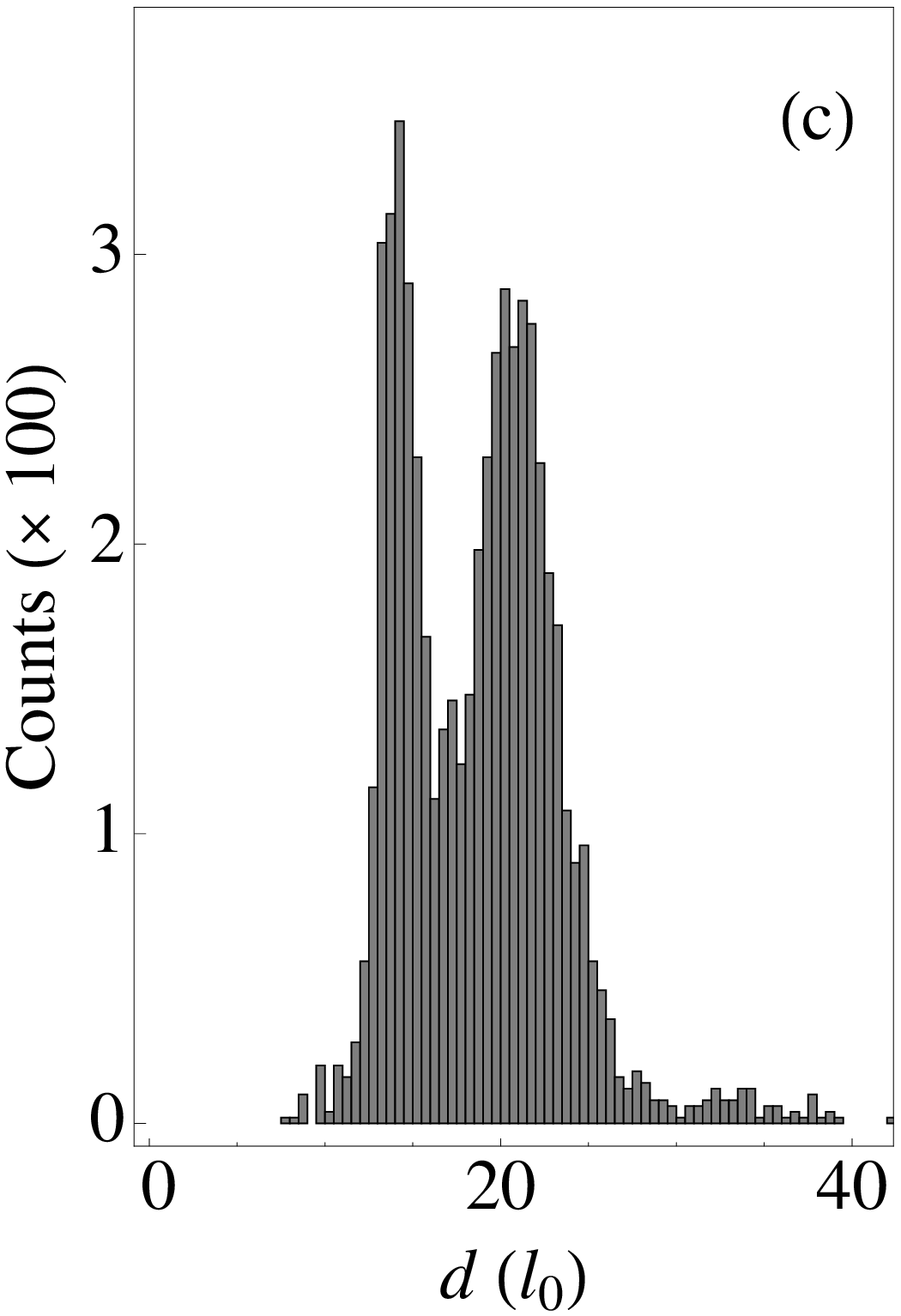}
\caption{Island distance distribution corresponding to Fig.~\ref{figislands}} \label{figdistrib}
\end{figure}

\begin{figure} \centering
\includegraphics[width=0.35\textwidth]{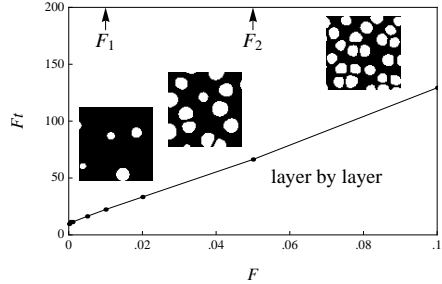}
\caption{The kinetic phase diagram as function of the deposited height $h_d \pe F t$ and 
the flux. For a given flux, growth proceeds mainly layer by layer 
up to the dynamical critical height $h_s$ (solid line). Above, for $F \ppp F_1$, surface diffusion 
is efficient and an annealing-like ripening proceeds. For $F_1 \ppp F \ppp F_2$, a non-conventional 
coarsening is at stake where the island density decreases faster than linearly. Finally, for 
$F \ppg F_2$, ripening is frozen by the deposition growth, the island density 
stays constant and highly correlated islands tend to form ridge-like structures. } \label{figdiagphas}
\end{figure}

The different nonlinear regimes are moreover characterized by different spatial organization, see
Fig.~\ref{figislands}. At low flux, the islands do not exhibit strong spatial correlations, similarly 
to the annealing case \cite{jnFrisVerg07}, while for higher flux, island decimation, driven by surface 
diffusion, leads to the emergence of island clusters. These clusters, absent in the annealing case, 
involve more and more islands as $F$ increases (Fig.~\ref{figislands}). 
To quantify these correlations, we first assign an area mass center to each island and construct a Delaunay
triangulation for this set of points. The typical nearest neighbor distance histograms in the nonlinear regimes 
are then plotted in Fig.~\ref{figdistrib}. At low flux, the histogram displays a broad bell shape with a 
modulation reminiscent of the instability initial stages, where, beside the first peak near the initial 
wavelength $\lambda \! \simeq \! 20 \lz$, the other peaks arise as coarsening is fully developed. 
For intermediate flux, the histogram is significantly narrowed with a main peak near $\lambda$, 
and other peaks which positions increase slightly with time as island decimation occurs.
In the steady high flux regime, Fig.~\ref{figdistrib}c), ordering is maximal and is 
described by a narrow distance distribution with two peaks related to the typical distance between 
two islands in a cluster and between clusters. 
The first peak arises at a position significantly lower than the instability wavelength $\lambda$, signaling a 
narrowing of the distance between mass centers during clustering at high flux, while the second 
peak sticks at $\lambda$. In this case, a particular ordering is observed in Fig.~\ref{figislands}c)  
as islands self-organize in ridge-like patterns which result from elastic interactions and stabilization of the 
nonlinear regime by the flux.  Finally, note
that the histograms of the Voronoi tessellation do not give relevant information here, 
contrarily to systems with island nucleation. 
These results are summarized by the kinetic phase diagram in Fig.~\ref{figdiagphas}.

As a conclusion, we studied the influence of a deposition flux on the growth dynamics 
leading to island formation. We used a continuum description of the elastic instability 
accounting for wetting and nonlocal elasticity effects where the flux arises 
in a nontrivial way. We found three different nonlinear regimes, depending on the flux 
and characterized by different spatial ordering and dynamics. At low flux, an 
annealing-like dynamics is at stake. 
Increasing the flux, we find spatial correlations where dots gather in clusters,  
together with a non-conventional ripening. For high flux, ripening is frozen as surface diffusion 
effects are inhibited by deposition growth and the dot density is  maximum. 
The nearest neighbor distance distribution exhibits a first peak linked to 
the distance between two islands in a cluster which decreases, and a second one related to 
the cluster distance. These results indicate a way to tune different 
spatial ordering and are calling for experimental examination.


\begin{thebibliography}{21}
\expandafter\ifx\csname natexlab\endcsname\relax\def\natexlab#1{#1}\fi
\expandafter\ifx\csname bibnamefont\endcsname\relax
  \def\bibnamefont#1{#1}\fi
\expandafter\ifx\csname bibfnamefont\endcsname\relax
  \def\bibfnamefont#1{#1}\fi
\expandafter\ifx\csname citenamefont\endcsname\relax
  \def\citenamefont#1{#1}\fi
\expandafter\ifx\csname url\endcsname\relax
  \def\url#1{\texttt{#1}}\fi
\expandafter\ifx\csname urlprefix\endcsname\relax\def\urlprefix{URL }\fi
\providecommand{\bibinfo}[2]{#2}
\providecommand{\eprint}[2][]{\url{#2}}

\bibitem[{\citenamefont{Shchukin et~al.}(2003)\citenamefont{Shchukin,
  Ledentsov, and Bimberg}}]{Shchukin}
\bibinfo{author}{\bibfnamefont{V.}~\bibnamefont{Shchukin}},
  \bibinfo{author}{\bibfnamefont{N.}~\bibnamefont{Ledentsov}},
  \bibnamefont{and} \bibinfo{author}{\bibfnamefont{D.}~\bibnamefont{Bimberg}},
  \emph{\bibinfo{title}{Epitaxy of Nanostructures}}
  (\bibinfo{publisher}{Springer}, \bibinfo{year}{2003})\bibinfo{note}{; V. A.
  Shchukin and D. Bimberg, Rev. Mod. Phys., \textbf{71}, 1125 (1999).}

\bibitem[{\citenamefont{Stangl et~al.}(2004)\citenamefont{Stangl, Holy, and
  Bauer}}]{Stangl04}
\bibinfo{author}{\bibfnamefont{J.}~\bibnamefont{Stangl}},
  \bibinfo{author}{\bibfnamefont{V.}~\bibnamefont{Holy}}, \bibnamefont{and}
  \bibinfo{author}{\bibfnamefont{G.}~\bibnamefont{Bauer}},
  \bibinfo{journal}{Rev. Mod. Phys.} \textbf{\bibinfo{volume}{76}},
  \bibinfo{pages}{725} (\bibinfo{year}{2004}).

\bibitem[{\citenamefont{Teichert}(2002)}]{Teichert02}
\bibinfo{author}{\bibfnamefont{C.}~\bibnamefont{Teichert}},
  \bibinfo{journal}{Physics Report} \textbf{\bibinfo{volume}{365}},
  \bibinfo{pages}{335} (\bibinfo{year}{2002}).

\bibitem[{\citenamefont{Pimpinelli and Villain}(1998)}]{Pimpinelli98}
\bibinfo{author}{\bibfnamefont{A.}~\bibnamefont{Pimpinelli}} \bibnamefont{and}
  \bibinfo{author}{\bibfnamefont{J.}~\bibnamefont{Villain}},
  \emph{\bibinfo{title}{Physics of Crystal Growth}}
  (\bibinfo{publisher}{Cambdrige University Press}, \bibinfo{year}{1998}).

\bibitem[{\citenamefont{Barabasi and Stanley}(1995)}]{BaraStan}
\bibinfo{author}{\bibfnamefont{A.-L.} \bibnamefont{Barabasi}} \bibnamefont{and}
  \bibinfo{author}{\bibfnamefont{H.~E.} \bibnamefont{Stanley}},
  \emph{\bibinfo{title}{Fractal concepts in surface growth}}
  (\bibinfo{publisher}{Cambridge University Press}, \bibinfo{year}{1995}).

\bibitem[{\citenamefont{Vvedensky}(in press)}]{Vved08}
\bibinfo{author}{\bibfnamefont{D.~D.} \bibnamefont{Vvedensky}},
  \emph{\bibinfo{title}{Frontiers in Nanoscience and Nanotechnology}}
  (\bibinfo{publisher}{Oxford University Press, Oxford, England},
  \bibinfo{year}{in press}), chap. \bibinfo{chapter}{Quantum Dots:
  Self-organized and self-limiting structures}.

\bibitem[{\citenamefont{Sutter and Lagally}(2000)}]{SuttLaga00}
\bibinfo{author}{\bibfnamefont{P.}~\bibnamefont{Sutter}} \bibnamefont{and}
  \bibinfo{author}{\bibfnamefont{M.~G.} \bibnamefont{Lagally}},
  \bibinfo{journal}{Phys. Rev. Lett.} \textbf{\bibinfo{volume}{84}},
  \bibinfo{pages}{4637} (\bibinfo{year}{2000}).

\bibitem[{\citenamefont{Tromp et~al.}(2000)\citenamefont{Tromp, Ross, and
  Reuter}}]{TromRoss00}
\bibinfo{author}{\bibfnamefont{R.~M.} \bibnamefont{Tromp}},
  \bibinfo{author}{\bibfnamefont{F.~M.} \bibnamefont{Ross}}, \bibnamefont{and}
  \bibinfo{author}{\bibfnamefont{M.~C.} \bibnamefont{Reuter}},
  \bibinfo{journal}{Phys. Rev. Lett.} \textbf{\bibinfo{volume}{84}},
  \bibinfo{pages}{4641} (\bibinfo{year}{2000}).

\bibitem[{\citenamefont{Berbezier et~al.}(2002)\citenamefont{Berbezier, Ronda,
  and Portavoce}}]{BerbRond02}
\bibinfo{author}{\bibfnamefont{I.}~\bibnamefont{Berbezier}},
  \bibinfo{author}{\bibfnamefont{A.}~\bibnamefont{Ronda}}, \bibnamefont{and}
  \bibinfo{author}{\bibfnamefont{A.}~\bibnamefont{Portavoce}},
  \bibinfo{journal}{J. Phys. : Condens. Matter} \textbf{\bibinfo{volume}{14}},
  \bibinfo{pages}{8283} (\bibinfo{year}{2002}).

\bibitem[{\citenamefont{Floro et~al.}(2000)\citenamefont{Floro, Sinclair,
  Chason, Freund, Twesten, Hwang, and Lucadamo}}]{Flor00}
\bibinfo{author}{\bibfnamefont{J.~A.} \bibnamefont{Floro}},
  \bibinfo{author}{\bibfnamefont{M.~B.} \bibnamefont{Sinclair}},
  \bibinfo{author}{\bibfnamefont{E.}~\bibnamefont{Chason}},
  \bibinfo{author}{\bibfnamefont{L.~B.} \bibnamefont{Freund}},
  \bibinfo{author}{\bibfnamefont{R.~D.} \bibnamefont{Twesten}},
  \bibinfo{author}{\bibfnamefont{R.~Q.} \bibnamefont{Hwang}}, \bibnamefont{and}
  \bibinfo{author}{\bibfnamefont{G.~A.} \bibnamefont{Lucadamo}},
  \bibinfo{journal}{Phys. Rev. Lett.} \textbf{\bibinfo{volume}{84}},
  \bibinfo{pages}{701} (\bibinfo{year}{2000}).

\bibitem[{\citenamefont{Gao and Nix}(1999)}]{GaoNix99}
\bibinfo{author}{\bibfnamefont{H.}~\bibnamefont{Gao}} \bibnamefont{and}
  \bibinfo{author}{\bibfnamefont{W.}~\bibnamefont{Nix}},
  \bibinfo{journal}{Annu. Rev. Mater. Sci} \textbf{\bibinfo{volume}{29}},
  \bibinfo{pages}{173} (\bibinfo{year}{1999}).

\bibitem[{\citenamefont{Spencer et~al.}(1991)\citenamefont{Spencer, Voorhees,
  and Davis}}]{SpenVoor91}
\bibinfo{author}{\bibfnamefont{B.~J.} \bibnamefont{Spencer}},
  \bibinfo{author}{\bibfnamefont{P.~W.} \bibnamefont{Voorhees}},
  \bibnamefont{and} \bibinfo{author}{\bibfnamefont{S.~H.} \bibnamefont{Davis}},
  \bibinfo{journal}{Phys. Rev. Lett.} \textbf{\bibinfo{volume}{67}},
  \bibinfo{pages}{3696} (\bibinfo{year}{1991}).

\bibitem[{\citenamefont{Spencer et~al.}(1993)\citenamefont{Spencer, Davis, and
  Voorhees}}]{Spencer93}
\bibinfo{author}{\bibfnamefont{B.~J.} \bibnamefont{Spencer}},
  \bibinfo{author}{\bibfnamefont{S.~H.} \bibnamefont{Davis}}, \bibnamefont{and}
  \bibinfo{author}{\bibfnamefont{P.~W.} \bibnamefont{Voorhees}},
  \bibinfo{journal}{Phys. Rev. B} \textbf{\bibinfo{volume}{47}},
  \bibinfo{pages}{9760} (\bibinfo{year}{1993}).

\bibitem[{\citenamefont{Xiang and E}(2002)}]{XianE02}
\bibinfo{author}{\bibfnamefont{Y.}~\bibnamefont{Xiang}} \bibnamefont{and}
  \bibinfo{author}{\bibfnamefont{W.}~\bibnamefont{E}}, \bibinfo{journal}{J.
  Appl. Phys.} \textbf{\bibinfo{volume}{91}}, \bibinfo{pages}{9414}
  (\bibinfo{year}{2002}).

\bibitem[{\citenamefont{Golovin et~al.}(2003)\citenamefont{Golovin, Davis, and
  Voorhees}}]{Golovin03}
\bibinfo{author}{\bibfnamefont{A.~A.} \bibnamefont{Golovin}},
  \bibinfo{author}{\bibfnamefont{S.~H.} \bibnamefont{Davis}}, \bibnamefont{and}
  \bibinfo{author}{\bibfnamefont{P.~W.} \bibnamefont{Voorhees}},
  \bibinfo{journal}{Phys. Rev. E} \textbf{\bibinfo{volume}{68}},
  \bibinfo{pages}{056203} (\bibinfo{year}{2003}).

\bibitem[{\citenamefont{Aqua et~al.}(2007)\citenamefont{Aqua, Frisch, and
  Verga}}]{jnFrisVerg07}
\bibinfo{author}{\bibfnamefont{J.-N.} \bibnamefont{Aqua}},
  \bibinfo{author}{\bibfnamefont{T.}~\bibnamefont{Frisch}}, \bibnamefont{and}
  \bibinfo{author}{\bibfnamefont{A.}~\bibnamefont{Verga}},
  \bibinfo{journal}{Phys. Rev. B} \textbf{\bibinfo{volume}{76}},
  \bibinfo{pages}{165319} (\bibinfo{year}{2007}).

\bibitem[{\citenamefont{{Levine} et~al.}(2007)\citenamefont{{Levine},
  {Golovin}, {Davis}, and {Voorhees}}}]{Levine07}
\bibinfo{author}{\bibfnamefont{M.~S.} \bibnamefont{{Levine}}},
  \bibinfo{author}{\bibfnamefont{A.~A.} \bibnamefont{{Golovin}}},
  \bibinfo{author}{\bibfnamefont{S.~H.} \bibnamefont{{Davis}}},
  \bibnamefont{and} \bibinfo{author}{\bibfnamefont{P.~W.}
  \bibnamefont{{Voorhees}}}, \bibinfo{journal}{Phy. Rev. B}
  \textbf{\bibinfo{volume}{75}}, \bibinfo{pages}{205312}
  (\bibinfo{year}{2007}).

\bibitem[{\citenamefont{Spencer}(1999)}]{Spen99}
\bibinfo{author}{\bibfnamefont{B.~J.} \bibnamefont{Spencer}},
  \bibinfo{journal}{Phys. Rev. B} \textbf{\bibinfo{volume}{59}},
  \bibinfo{pages}{2011} (\bibinfo{year}{1999}).

\bibitem[{\citenamefont{Tu and Tersoff}(2004)}]{TuTers04}
\bibinfo{author}{\bibfnamefont{Y.}~\bibnamefont{Tu}} \bibnamefont{and}
  \bibinfo{author}{\bibfnamefont{J.}~\bibnamefont{Tersoff}},
  \bibinfo{journal}{Phys. Rev. Lett.} \textbf{\bibinfo{volume}{93}},
  \bibinfo{pages}{216101} (\bibinfo{year}{2004}).

\bibitem[{\citenamefont{Tersoff et~al.}(2002)\citenamefont{Tersoff, Spencer,
  Rastelli, and von K\"anel}}]{TersSpen02}
\bibinfo{author}{\bibfnamefont{J.}~\bibnamefont{Tersoff}},
  \bibinfo{author}{\bibfnamefont{B.~J.} \bibnamefont{Spencer}},
  \bibinfo{author}{\bibfnamefont{A.}~\bibnamefont{Rastelli}}, \bibnamefont{and}
  \bibinfo{author}{\bibfnamefont{H.}~\bibnamefont{von K\"anel}},
  \bibinfo{journal}{Phys. Rev. Lett.} \textbf{\bibinfo{volume}{89}},
  \bibinfo{pages}{196104} (\bibinfo{year}{2002}).

\bibitem[{\citenamefont{Beck et~al.}(2004)\citenamefont{Beck, van~de Walle, and
  Asta}}]{abinitio}
\bibinfo{author}{\bibfnamefont{M.~J.} \bibnamefont{Beck}},
  \bibinfo{author}{\bibfnamefont{A.}~\bibnamefont{van~de Walle}},
  \bibnamefont{and} \bibinfo{author}{\bibfnamefont{M.}~\bibnamefont{Asta}},
  \bibinfo{journal}{Phys. Rev. B} \textbf{\bibinfo{volume}{70}},
  \bibinfo{pages}{205337} (\bibinfo{year}{2004})\bibinfo{note}{; G.-H. Lu and
  F. Liu, Phys. Rev. Lett. \textbf{94}, 176103 (2005)}.

\end{thebibliography}

\end{document}